# A Human-Centred Architecture for Large Language Models-Cognitive Assistants in Manufacturing within Quality Management Systems

Marcos Galdino[a]*, Johanna Grahl[a], Tobias Hamann[a], Anas Abdelrazeq[a], Ingrid Isenhardt[a]

[a]*IQS Intelligence in Quality Sensing, Laboratory for Machine Tools and Production Engineering (WZL) of RWTH Aachen University, RWTH Aachen University, Aachen, Germany*

\* Corresponding author. Tel.: +49-178-67-32321; *E-mail address:* marcos.galdino@wzl-iqs.rwth-aachen.de

**Abstract**

Large Language Models-Cognitive Assistants (LLM-CAs) can enhance Quality Management Systems (QMS) in manufacturing, fostering continuous process improvement and knowledge management. However, there is no human-centred software architecture focused on QMS that enables the integration of LLM-CAs into manufacturing in the current literature. This study addresses this gap by designing a component-based architecture considering requirement analysis and software development process. Validation was conducted via iterative expert focus groups. The proposed architecture ensures flexibility, scalability, modularity, and work augmentation within QMS. Moreover, it paves the way for its operationalization with industrial partners, showcasing its potential for advancing manufacturing processes.

## 1. Introduction

Large Language Models-Cognitive Assistants (LLM-CAs) can contribute to improving operational performance in manufacturing within Quality Management Systems (QMS) [1]. In this context, LLM-CAs must also fulfill QMS requirements, once being part of it.

LLM-CAs are artificial intelligence-driven systems that process and generate human-like text often used to support decision-making, automation, and communication. They bridge humans and technology, augmenting cognitive tasks and operational processes, e.g., knowledge sharing [2,3]. However, their implementation in manufacturing still faces challenges, e.g., hallucination [4] and lack of data privacy and security [5]. QMS, as per ISO 9001:2015, provide a structured framework to ensure consistent product quality, regulatory compliance, and continuous improvement [6]. One of the pivotal aspects of QMS is to ensure that knowledge is available to achieve conformity of products and services [7].

Although many authors address the development of LLM-CAs, none provides a holistic software architecture solution focused on their implementation for manufacturing within QMS. Moreover, the current solutions do not combine fine tuning and retrieval augmented generation (RAG), expertise-based hierarchical update of knowledge bases, compliance, fact, and jailbreak checks. Such requirements have been reported in the mentioned domain through different studies [1].

Besides, a central point in this study is that QMS require a structured document management approach, once documentation is subjected to audits. The implementation of LLM-CAs within QMS assumes, therefore, the existence of an extensive, versioned data corpora, e.g., work instructions, best practices, and machine manuals. Combining LLM-CAs and QMS could guarantee a smooth integration of LLM-CAs in manufacturing as it would benefit from well stablished industrial standards. Therefore, we believe that a successful implementation of LLM-CAs in manufacturing builds upon the principles of QMS. On that basis, this study aims to present an LLM-CA software architecture for QMS that enables efficient and reliable human-centred knowledge management and continuous improvement. To achieve that, the following research question will be answered:

*How can a system architecture for human-centred LLM-CAs in manufacturing within QMS be designed?*

To answer this question a requirement analysis for LLM-CAs in manufacturing within QMS carried out by Galdino et al. [1] is considered followed by software development process and validation through focus groups.

This paper is organised as follows: Chapter 2 provides the theoretical background regarding LLM-CAs, QMS, and current solutions for LLM-CAs. Chapter 3 outlines the methods. Chapter 4 presents the results, providing detailed depiction and explanation of the developed architecture. Finally, Chapter 5 concludes the paper contextualizing the architecture within QMS, summarizing the key findings and providing an outlook for future research.

## 2. Theoretical Background

*2.1. Large Language Models-Cognitive Assistants*

Large language models (LLMs) transformed natural language processing (NLP) by developing advanced architectures and vast datasets to generate human-like communication [8]. This development was enabled by the introduction of the machine-learning Transformer architecture by Vaswani et al. [9]. The Transformer architecture utilizes an attention mechanism that allows for parallel processing instead of sequential computation [9].

For instance, Fan et al. [10] explored the potential of LLM agents for industrial robotics. In robotic task planning, they achieved a success rate of 82%, using a GPT-4 agent. Kernan Freire et al. [11] proposed an LLM-CA in a factory setting aiming at answering queries and sharing of knowledge. In this setting, GPT-4 demonstrated superior performance, closely followed by open-source LLMs, e.g., StableBeluga2 and Mixtral 8x7B.

LLM-CAs present a potential disruptive impact in manufacturing. On the one hand, their deployment still faces various technical challenges, e.g., reasoning errors [3]. On the other hand, human and organisational challenges, e.g., artificial intelligence avoidance and lack of skilled workforce must also be considered [12]. On that basis, the deployment of LLM-CAs depend on their effective interplay with the subsystems "humans, technology, and organisation" [1].

*2.2. Quality Management Systems (ISO 9001:2015)*

QMS, as defined by ISO 9001:2015, establish a structured approach to ensuring consistent quality and organizational excellence [13]. The current version of ISO 9001:2015 emphasizes risk-based thinking, leadership commitment, stakeholder engagement, process management, and continuous improvement [14]. According to ISO 9001:2015 [7], a QMS defines objectives, manages resources and processes, optimizes decision-making, and identifies actions to address intended and unintended consequences. This approach guides organizations in developing, implementing, and improving their QMS to align with strategic goals and customer requirements [13].

Empirical studies confirm the benefits of QMS adoption. For instance, Solomon et al. [15] found that QMS improved efficiency and sustainability in the electricity sector by fostering quality culture and process optimization. More recently, Mustroph and Rinderle-Ma [16] conceptualized a software as a service-based QMS aligned with the EU Artificial Intelligence Act (EU AI Act) to monitor artificial intelligence system design, quality, and risk management.

Despite its broad applicability across industries, QMS implementation presents challenges, e.g., innovation and workforce engagement, cybersecurity and data protection, and resistance to change and leadership issues [17]. However, as manufacturing advances, the adoption of QMS remains crucial for sustaining competitiveness and ensuring quality-driven technological integration [18].

*2.3. Current solutions for LLM-CAs*

The current literature on LLM-CAs presents various solutions and architectures for their implementation.

Kernan Freire [2] implemented an LLM-CA in a factory setting for knowledge sharing. His LLM-CA dynamically updates the knowledge base with operator knowledge by means of a knowledge graph and dialectic interaction. His results highlight an overall positive employee perception of the LLM-CA in a detergent factory. Employees reported various benefits, e.g., improved knowledge sharing and problem-solving. However, his technical approach does not comprise mechanisms for safeguarding knowledge quality, approving procedures, compliance check, and guaranteeing that the knowledge base is always up to date.

Bucaioni et al. [19] proposed a reference architecture for the integration of LLMs into software systems. First, they identified architectural concerns regarding LLMs, e.g., data handling, performance, scalability, among others. Second, based on the concerns, they developed a modular architecture, which aims at providing a general approach for LLM integration. Nevertheless, there is no elucidation of user feedback mechanisms. In addition, their architecture does not comprise access control that regulates user interaction within the architecture. Their solution could be transferred to QMS. However, missing specific requirements, e.g., documentation management, presents potential further development.

The previous two examples demonstrate that research on LLM integration has advanced in the last years. However, the synergy between LLMs as an emerging technology and the specific requirements of QMS is still missing.

## 3. Methods

The proposed architecture has been developed using software development process, utilizing Domain-Driven Design (DDD). The result is a microservice-based multi-agent system (MAS).

Software development is a conceptual, iterative process. It typically consists of the phases: requirement analysis, design, implementation and testing [20,21]. Our work in this paper covers the requirement analysis and design phase.

Requirements analysis marks the beginning of the development process and involves the examination of the problem [21]. Requirements extracted in this phase form the basis for the system to be built. They are divided into functional (FR) and non-functional (NFR) requirements. FRs define the functionality of the system, e.g., the system shall incorporate human feedback into its knowledge base, whereas NFRs define the system qualities and properties, e.g., the system should be scalable and adaptable [22]. Table 1 presents the fundamental requirements of our system, which are the basis of our cognitive assistant system design. These requirements originate from an earlier work [1] and represent the technically operationalizable requirements. On that basis, more detailed requirements were defined, which were then divided into FR and NFR and.

Table 1: Requirements operationalized into the LLM-CA software architecture (Adapted from Galdino et al. [1])

| ID | Requirement |
|----|-------------|
| 1  | Enable artificial intelligence trustworthiness |
| 2  | Improve resistance to adversarial input |
| 3  | Enhance adaptability and scalability |
| 4  | Ensure reliable model performance |
| 5  | Guarantee effective memory usage |
| 6  | Ensure accurate language output |
| 7  | Ensure robust language understanding |
| 8  | Include human-in-the-loop |
| 9  | Design user-centric cognitive assistants |
| 10 | Enable effective communication-driven decision-making |
| 11 | Develop a skilled workforce |
| 12 | Adapt to factory environments |
| 13 | Integrate industry-specific knowledge |
| 14 | Ensure compliance integrity |

In the design phase, a solution to the problem, i.e., identified research gap, defined by the requirements is created [21]. This paper addresses solely the software architecture, defined as the high-level design of the system architecture and the result of made design decisions [23,20,21]. The architecture is derived from the FRs and NFRs in an iterative design process that includes its design, evaluation and modification [23,24,20]. The design process is determined by reusability, e.g. architectural patterns, methods of overcoming the gap between architecture and requirements, e.g., DDD, and intuition, e.g., architect experience [23,24].

High-level design decisions relate to FRs and are used along the further development of the architecture. Our design decisions are:

- Usage of Retrieval-Augmented Generation (RAG) pattern and LLM domain adapters:
  RAG was chosen as it allows to add real-time and updated knowledge to an LLM [25,26]. Adapters allow a domain specific parametric efficient fine-tuning of an LLM [27,28]. RAG systems in combination with fine-tuning outperform RAG systems without any fine-tuning [29]. The combination of both approaches shall increase the domain specific information in the output of the LLM.
- Usage of a conversational agent:
  The textual and linguistic use of the cognitive assistant (CA) should be enabled by a conversational agent based on the concept presented in [30].
- Generalization of Security Checks:
  Security checks are required to counteract potentially harmful content, such as incorporated user feedback. This is due to the fact that RAG systems are susceptible to malicious content in their database [31].

DDD approaches the development of complex software by focusing on one core domain and defining multiple bounded contexts with their own ubiquitous languages [32]. Bounded contexts describe the context within a model where boundaries are clearly defined [33]. To derive bounded contexts from FRs the English informal strategy, firstly introduced by Abbott [34], was used. This strategy systematically analyzes textual description of requirements by mapping nouns to possible objects, verbs to possible procedures and adjectives to possible properties. The extracted words are then used to create a domain model. A domain model characterizes the problem space [35]. Despite this analysis, real-world knowledge and domain expertise is still required to create such a model, as it is not enough to solely rely on the analysis [34]. Having a unified domain model for an entire system is not suitable. Thus, domain partitioning into multiple bounded contexts, which should be unified, is necessary [32]. On that basis, the following bounded contexts were derived through the analysis of the FRs combined with the high-level design decisions and the resulting partitioning of the domain model:
*Chat, Retriever, LLM, Knowledge base, Feedback, Conversational Agent, User.*

To further design the software architecture, architectural design options must be selected and properties and relationships of software components must be defined [23].

The discovered bounded contexts were used as the foundation for the design of the software architecture. Based on the properties of bounded contexts, suggestions exist to implement each bounded context as a microservice, creating a modular microservice architecture [36]. Microservices are cohesive, independent processes that are loosely coupled, independently deployable, and communicate via messages [37–39].

A MAS decomposes the system into multiple autonomous entities, such as agents or LLM-based agents, that work together to achieve the requirements of a system [40,21]. LLM-based agents enhance the capabilities of LLMs by integrating interactive problem-solving functions, so called function calling. They have expanded the range of problems addressable by LLMs, including those that require external information through interaction with software tools [41,40].

The resulting architecture design is a microservice-based MAS, where the architectural components are microservices that simultaneously function as agents. Such architecture is flexible, scalable, and interchangeable.

After having developed a first version of the architecture, the next phase consisted of its design evaluation [23] by focus groups. Focus group is a qualitative research method that gathers a small, diverse group of individuals to discuss specific topics, aiming to explore their attitudes, perceptions, beliefs, and opinions through guided interaction. [42] Seven domain experts participated in two online focus groups, apart one month from each other, to assess aspects of the proposed architecture. The domain experts stemmed from academia and brought different levels of experience in research, software development, QMS, and LLMs. In the first focus group, we presented the architecture to them who focused on identifying strengths, limitations, and areas for improvement, ensuring alignment with user needs and industry standards. Based on their feedback, we improved the architecture, which was again presented to the same seven domain experts in the second focus group. They analysed and provided feedback to the second version, which was updated based on the discussion. At the end of the second focus group, a final version of the architecture was presented and accordingly accepted by the participants.

## 4. Results

The following chapter presents the resulting design of the architecture, shown in Figure 1. A brief overview of the architecture, the relationships between its components, and their descriptions is provided.

Collaboration and communication are orchestrated by architectural component *ChatController*. It also serves as the interface to the user interface (UI). Each core functionality is represented by a controller, which manages its interactions with the other controllers. The textual and vocal usage is orchestrated by *ConversationalAgent Controller* through the interaction between the architecture components *ConversationalAgent*, *LLMAgent* and *RAGRetrieval*. To incorporate user feedback into the system, *FeedbackEvaluation*

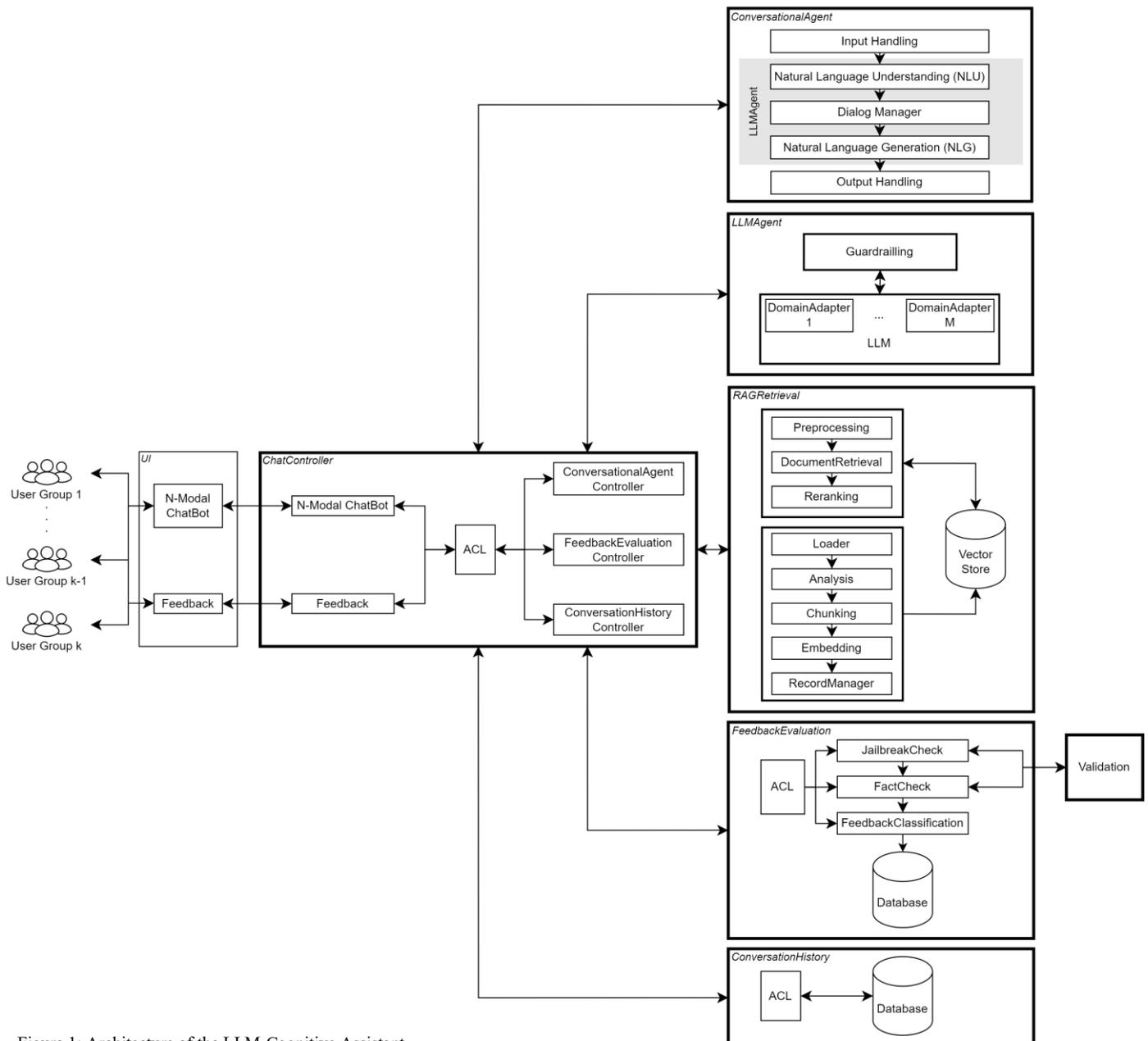

Figure 1: Architecture of the LLM-Cognitive Assistant

*Controller* manages the interaction between *FeedbackEvaluation* and *RAGRetrieval*.

The controller for *ConversationHistory* is solely responsible for providing this component. Users may be assigned to several user groups, with different authorizations, which are managed through an Access Control List (ACL).

The component *ConversationalAgent* enables verbal and textual interaction with the cognitive assistant, ensuring a multimodal system with at least two modalities. It is based on the conversational agent architecture proposed by [30]. Component *Input Handling* converts supported input formats into a textual representation, whereas *Output Handling* converts the response of the system into the needed output format, i.e., text or voice. The components Dialog Manager, Natural Language Understanding (NLU) and Natural Language Generation (NLG) are realized by the *LLMAgent*, as indicated by the grey background of the components in the architecture diagram. Architectural component *LLMAgent* is an LLM-based agent, which comprises two components: Guardrailling and LLM. Guardrailling is a safety measurement to regulate user interactions with LLMs with the purpose of ensuring that the LLM adheres to ethical and organization principles, e.g., compliance guidelines based on government regulations and employee equality policies. It is realized by identifying harmful content in user prompts and model responses [43,44]. The second component contains an LLM with multiple domain adapters. Adapters allow to fine tune a general purpose LLM for a domain specific task, while still using the capabilities of an LLM without requiring extensive computational resources [27,45]. An LLM chosen for this component needs to provide the functionality of function calling to enable the workflow of an LLM-based agent. Function calling further enables users to use mathematical functions needed for process-specific calculations and process data analysis that would serve as a first step to empower users to perform continuous improvement, e.g., optimization of product setup time.

Architectural component *RAGRetrieval* manages and retrieves documents necessary for injecting an LLM with up-to-date knowledge. Documents can be loaded from a defined source, e.g., file system, as well as directly updated, e.g., by a feedback ticket. Documents are first analysed, considering aspects such as layout and table structure, and subsequently unified into a common format based on the concept presented in [46]. Afterwards, they are chunked, embedded and stored in a vector store. A second component utilizes the retrieval of document chunks for a given query.

Architecture component *FeedbackEvaluation* enables the functionality to continuously improve the systems knowledge through users. Users can either flag an answer as "insufficient" for an incorrect answer, or as "extend" for a partially complete answer, creating a feedback ticket. Specific user groups are allowed to rewrite such answers, e.g., user group k-1 (supervisors), and extend a feedback ticket with additional documents or update existing documents. Feedback tickets undergo two security checks before being integrated into the vector store. They consist of a jailbreak check and a fact check. The jailbreak checker verifies whether the feedback ticket does not contain adversarial input that could change the behavior of the LLM, e.g. jamming attacks [31]. The fact checker ensures that injected knowledge is within the scope of context provided by the feedback ticket. It is possible for both checks to be performed by humans, by an LLM, or by their combination.

To enable continuous learning, feedback tickets are stored and can be retrieved by specific user groups, e.g. user group k at the managerial level. These tickets can provide anonymous insights into potential user knowledge gaps, e.g., lack of data analytics skills, and system response quality, e.g., rate of incomplete answers. Architectural component *ConversationHistory* stores and provides all user's conversations for resumption.

## 5. Conclusions

The findings of this study highlight the potential of LLM-CAs to enhance operational performance in manufacturing within QMS. By integrating artificial intelligence-driven capabilities, LLM-CAs can facilitate decision-making, continuous improvement, and knowledge management while aligning with QMS requirements.

Existing LLM-CAs research does not focus on QMS, which can be fundamental for the integration of such a technology into manufacturing [1]. This research addresses this gap by proposing a holistic software architecture for LLM-CAs within QMS. The proposed solution integrates fine-tuning, RAG, hierarchical knowledge updates, and compliance mechanisms, ensuring efficient, auditable, and human-centred knowledge management and continuous improvement. On the one hand, the architecture enables employees to be key players within work activities, e.g., performing process improvements. On the other hand, it supports them by means of technological functionalities, e.g., data analysis, version control of documents, and compliance checks.

The limitation of this study lies on its theoretical development, assuming therefore its full functionality without previous testing. On that basis, our future research will operationalize and evaluate the proposed architecture based on real manufacturing use cases and user evaluation. Moreover, different available LLMs will be tested for performance. Therefore, practical results can lead to an update of the current proposed architecture that would refine it and enhance its applicability in manufacturing.

**Acknowledgements**

This research was developed in the context of the project ZUKIPRO [ZUK-1-0005], which is funded by the German Federal Ministry of Labour and Social Affairs (BMAS) and the European Union via the European Social Fund Plus (ESF Plus) as part of the 'Future Centres' programme.